\documentclass[sn-mathphys-num]{sn-jnl}% Math and Physical Sciences Numbered Reference Style 
\usepackage{graphicx}%
\usepackage{multirow}%
\usepackage{amsmath,amssymb,amsfonts}%
\usepackage{amsthm}%
\usepackage{mathrsfs}%
\usepackage[title]{appendix}%
\usepackage{xcolor}%
\usepackage{textcomp}%
\usepackage{manyfoot}%
\usepackage{booktabs}%
\usepackage{algorithm}%
\usepackage{algorithmicx}%
\usepackage{algpseudocode}%
\usepackage{listings}%

\usepackage{mathtools}
\usepackage{appendix}
\usepackage{hyperref}
\hypersetup{colorlinks=true,linkcolor=blue,urlcolor=blue,citecolor=blue}

\renewcommand{\thetable}{\arabic{table}}

\usepackage{makecell}
\newcommand{\tabthickness}{0.5mm}
\newcolumntype{?}[1]{!{\vrule width #1}}

\newcommand{\p}[1]{^{(#1)}}

\theoremstyle{thmstyleone}%
\theoremstyle{thmstyletwo}%

\theoremstyle{thmstylethree}%

\begin{document}

\title[Article Title]{A multiple-timing analysis of temporal ratcheting}

%%=============================================================%%
%% GivenName	-> \fnm{Joergen W.}
%% Particle	-> \spfx{van der} -> surname prefix
%% FamilyName	-> \sur{Ploeg}
%% Suffix	-> \sfx{IV}
%% \author*[1,2]{\fnm{Joergen W.} \spfx{van der} \sur{Ploeg} 
%%  \sfx{IV}}\email{iauthor@gmail.com}
%%=============================================================%%

\author*[1,2]{\fnm{Aref} \sur{Hashemi}}\email{aref.hashemi@nd.edu}
\author[3]{\fnm{Edward T.} \sur{Gilman}}
%% \equalcont{These authors contributed equally to this work.}
\author*[4]{\fnm{Aditya S.} \sur{Khair}}\email{akhair@andrew.cmu.edu}
%% \equalcont{These authors contributed equally to this work.}

\affil*[1]{\orgdiv{Department of Applied and Computational Mathematics and Statistics}, \orgname{University of Notre Dame}, \city{Notre Dame}, \state{IN}, \country{United States}}

\affil[2]{\orgdiv{Courant Institute}, \orgname{New York University}, \city{New York}, \state{NY}, \country{United States}}

\affil[3]{\orgdiv{Department of Mathematics}, \orgname{Rice University}, \city{Houston}, \state{TX}, \country{United States}}

\affil*[4]{\orgdiv{Department of Chemical Engineering}, \orgname{Carnegie Mellon University}, \city{Pittsburgh}, \state{PA}, \country{United States}}

%%==================================%%
%% Sample for unstructured abstract %%
%%==================================%%

\abstract{We develop a two-timing perturbation analysis to provide quantitative insights on the existence of temporal ratchets in an exemplary system of a particle moving in a tank of fluid in response to an external vibration of the tank. We consider two-mode vibrations with angular frequencies $\omega$ and $\alpha\omega$, where $\alpha$ is a rational number. If $\alpha$ is a ratio of odd and even integers (e.g., $\tfrac{2}{1},\,\tfrac{3}{2},\,\tfrac{4}{3}$), the system yields a net response: here, a nonzero time-average particle velocity. Our first-order perturbation solution predicts the existence of temporal ratchets for $\alpha=2$. Furthermore, we demonstrate, for a reduced model, that the temporal ratcheting effect for $\alpha=\tfrac{3}{2}$ and $\tfrac{4}{3}$ appears at the third-order perturbation solution. More importantly, we find closed form formulas for the magnitude and direction of the induced net velocities for these $\alpha$ values. On a broader scale, our methodology offers a new mathematical approach to study the complicated nature of temporal ratchets in physical systems.}


%%\pacs[JEL Classification]{D8, H51}

%%\pacs[MSC Classification]{35A01, 65L10, 65L12, 65L20, 65L70}

\maketitle

\section{Introduction}\label{sec1}
A zero time-average oscillatory excitation with certain \emph{broken time symmetries} can induce a net response, i.e., a nonzero time-average solution, in nonlinear dynamical systems \cite{Flach2000,Denisov2002,Reimann2002,Ustinov2004,Hanggi2009,Denisov2014,Cubero2016}. In particular, periodic excitations that are not shift-symmetric (or antiperiodic) can induce a net drift when acting on nonlinear systems \cite{Denisov2014,Freire2013}. Unlike the classical Feynman-Smoluchowski ratchet \cite{Kalinay2018FSRatchet}, where the symmetry is broken in the physical domain, these ``temporal ratchets'' are caused by time asymmetries. Previous theoretical and experimental studies have established the existence of temporal ratchets in point particles \cite{Flach2000,Yevtushenko2001,Denisov2002,Dukhin-Dukhin2005,deGennes2005JStatPhys,Cubero2010,Wickenbrock2011}, optical \cite{Schiavoni2003,Jones2004,Gommers2005,Gommers2006,Struck2012,Eckardt2017} and quantum \cite{Denisov2007a,Denisov2007b} lattices, and mechanical and microfluidic systems \cite{Vidybida1985,Eglin2006,Buguin2006EPJE,Fleishman2007,Aref2022PRE,Daniel2005Langmuir}. However, the underlying mechanism by which the temporal ratcheting is induced has remained relatively obscure. Notably, it is unclear as to what determines or, rather, how to predict the sign and magnitude of the net response, e.g., the direction and speed of net motion. Typically, the sign and magnitude of the net response are determined by a direct numerical integration of the governing equation. Here, in this study, we introduce multiple time-scale theory as a method to analyze temporal ratchets. Our approach shows a new way to obtain an analytical approximation to the ratcheting behavior and, thereby, provides a theoretical understanding of their origin. Importantly, we provide a new methodology to find closed-form formula for the magnitude and direction of the net motion in temporal ratchets.

\begin{figure}[t]
  \centering
  \setlength{\belowcaptionskip}{-10pt}
  \includegraphics{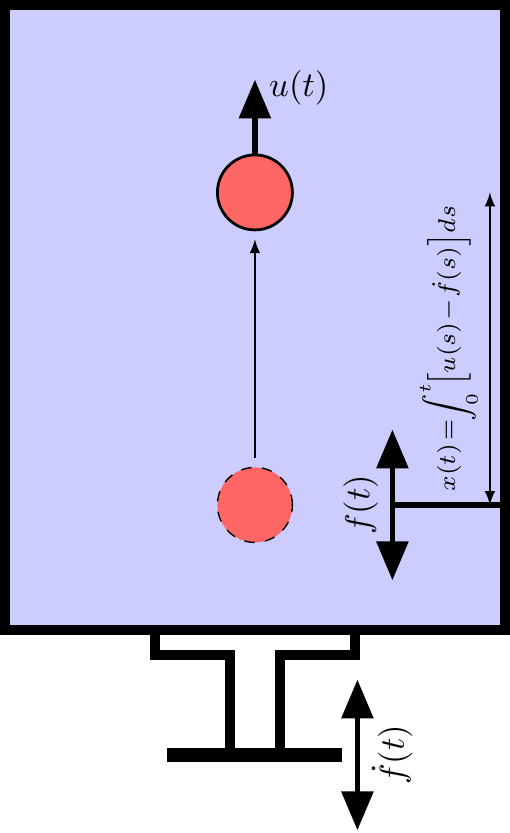}
  \caption{Schematic diagram of the problem. A particle moves at a velocity $u(t)$ in a large tank that is fully filled with liquid and vibrates vertically at a periodic velocity given by $\dot{f}(t)$. $x(t)$ denotes the vertical location of the particle relative to the tank. The tank vibrates up and down with a displacement $f(t)$ with respect to the initial location of the particle.}
  \label{fig:schematic}
\end{figure}

\section{Physical Problem}
As a model system, consider a particle immersed in a large (compared to the size of the particle) enclosed tank, which is filled with a liquid and vertically placed on top of a platform (see \autoref{fig:schematic}). The platform vibrates up and down with a periodic wave form $f(t)=\ell\tilde{f}(t)$, where $\ell$ is the nominal amplitude of oscillation, and $t$ is time. The Newton's second law for the particle motion can be expressed as
\begin{equation}
\label{eq:Newtonfm}    
m\dot{u}=-a\,(u-\dot{f})-b\,\mathrm{sgn}(u-\dot{f})(u-\dot{f})^2-Vg\Delta\rho,    
\end{equation}
where $m$ and $u$ are the particle mass and velocity, respectively; $\mathrm{sgn}(x)$ is the sign function; $a$ and $b$ are positive constants; $Vg\Delta\rho$ is the gravitational force on the particle; and overdot denotes differentiation with respect to time. The terms on the right hand side (RHS) of \eqref{eq:Newtonfm} correspond to linear drag/friction, a nonlinear drag, and gravitational forces, respectively.

Using $\omega^{-1}$ and $\ell\omega$ as, respectively, the time and velocity scales, and changing the dependent variable to $v=u-\dot{f}$ (with $v$ as the particle velocity relative to the tank), we find the dimensionless form of \eqref{eq:Newtonfm},
\begin{equation}
  \label{eq:Newtonfm-dl}
  \frac{d\tilde{v}}{d\tilde{t}}=-\frac{d^2\tilde{f}}{{d\tilde{t}}^2}-k\tilde{v}-\epsilon\,\mathrm{sgn}(\tilde{v})\tilde{v}^2-w,
\end{equation}
where $k=a/(m\omega)$, $\epsilon=b\ell/m$, and $w=Vg\Delta\rho/(m\ell\omega^2)$. To make further progress, we approximate the $\mathrm{sgn}(\tilde{v})$ function with Taylor series expansion of a smooth sigmoid function $\tanh(\tilde{v})$,
\begin{equation}
  \label{eq:Newtonfm-dl-sigmoid}
  \frac{d\tilde{v}}{d\tilde{t}}=-\frac{d^2\tilde{f}}{{d\tilde{t}}^2}-k\tilde{v}-\epsilon\left(\tilde{v}-\tfrac{1}{3}\tilde{v}^3+\,\cdots\right)\tilde{v}^2-w.
\end{equation}
While the approximation $\mathrm{sgn}(\tilde{v})\approx\tanh(\tilde{v})$ might be quantitatively inaccurate, it does capture the most important qualitative characteristics of the resistance term, such as being nonlinear, odd in $v$, and monotonic increasing. Finally, for $\mid\tilde{v}\mid\ll 1$, which physically implies that the particle moves at only a slightly different velocity than the tank, we truncate the series to obtain
\begin{equation}
  \label{eq:governing_full}
  \frac{dv}{dt}=-\frac{d^2f}{{dt}^2}-kv-\epsilon v^3-w,
\end{equation}
where we dropped the $\sim$ decoration for simplicity of the presentations. To close the problem, we assume the initial condition $v(t=0)=0$, i.e., the relative velocity is initially zero.

\begin{figure}[t]
  \centering
  \setlength{\belowcaptionskip}{-10pt}
  \includegraphics{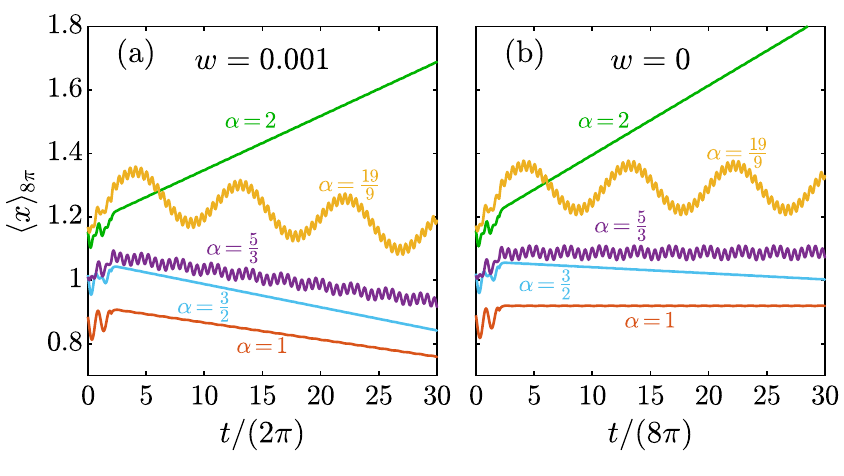}
  \caption{Representative numerical solution to \eqref{eq:governing_full}. Moving average (on time intervals of length $8\pi$) of the relative position of the particle, $x(t)=\int_0^tv(s)ds$, versus time for different $\alpha$ values with $w=0.001$ (a) and $w=0$ (b). Parameters: $c=1$, $k=1$, $\epsilon=0.25$.}
  \label{fig:x_vs_t_diffAlpha}
\end{figure}

To proceed, we assume a specific two-mode excitation
\begin{equation}
  \label{eq:f}
  f(t)=\tfrac{1}{2}\left[\sin(t)+c\,\sin(\alpha t)\right],  
\end{equation}
with $\alpha$ as a rational number, and $2\tau=2\pi/\mathrm{gcd}(1,\alpha)$, where $\gcd(1,\alpha)$ denotes the greatest common divisor of $1$ and $\alpha$. Without any loss of generality, we assume $\alpha\ge1$; any two-mode waveform with $\alpha<1$ can be expressed as another waveform with $\alpha\coloneqq1/\alpha$ by choosing $(\alpha\omega)^{-1}$ as the time scale. In \eqref{eq:f}, $c$ is assumed to be an $O(1)$ constant that represents the relative importance of the two modes of excitation.

\autoref{fig:x_vs_t_diffAlpha} shows representative numerical results for the particle position (relative to the tank), i.e., $x(t)=\int_0^tv(s)ds$, versus time for different values of $\alpha$, and with and without gravitational forces. (Note that we are not interested in cases where the system behavior is largely dominated by gravitational forces; so we present results for $w=0.001$ and $w=0$ only.) When $w\neq 0$ (\autoref{fig:x_vs_t_diffAlpha}(a)), one would expect the particle to follow, on average, the direction of the gravitational force. This is the case for $\alpha=1,\,\tfrac{3}{2},\,\tfrac{5}{3},\,\tfrac{19}{9}$; the particle moves on average downward despite oscillating. Perhaps the simplest intuition is that in the absence of the nonlinear term, one could simply find the average velocity as $\langle v\rangle=-w/k$. (Here, $\langle\chi\rangle$ denotes a time-average over the period $2\tau$: namely, $\langle \chi\rangle=\frac{1}{2\tau}\int_{t}^{t+2\tau}\chi ds$.) However, the system behaves qualitatively different for $\alpha=2$ and the particle moves upward on average, against gravity.

Further analysis indicates that there are two different phenomena governing the average motion of the particle. \autoref{fig:x_vs_t_diffAlpha}(b) shows the same results but for $w=0$. Here, the particle does not move, on average, for $\alpha=1,\,\tfrac{5}{3},\,\tfrac{19}{9}$ (the particle experiences an initial drift but then remains, on average, in the same location) but drifts downward and upward with a \emph{constant velocity} for $\alpha=\tfrac{3}{2}$ and $\alpha=2$, respectively. Note that the total resistance term in \eqref{eq:governing_full} is nonlinear and odd in $v$. The latter indicates that the system has no direction bias when $w=0$. Under such conditions, it can be proved that an antiperiodic $f(t)$, for which $f(t+\tau)=-f(t)$, invariably yields a zero net response, i.e., $\langle v\rangle=0$. However, if such a symmetry is broken, the system will, potentially, behave as a ratchet; see Hashemi et al. \cite{Aref2022PRE} for details. It is straightforward to show that if $\alpha$ can be expressed as a ratio of two odd integers (e.g., $\tfrac{1}{1},\,\tfrac{5}{3},\,\tfrac{19}{9}\dots$), $f(t)$ is antiperiodic, and hence, yields no net response \cite{Aref2022PRE}. This is not necessarily true when $f(t)$ is non-antiperiodic, e.g., for $\alpha=\tfrac{3}{2}$ and $2$; as shown in \autoref{fig:x_vs_t_diffAlpha}(b), the particle experiences a constant drift for these $\alpha$ values. In short, the two phenomena governing the net motion of the particle are (i) spatial ratcheting effect due to the presence of gravitational forces, and (ii) temporal ratcheting due to non-antiperiodic excitations. In fact, one can see the system behavior as a superposition of these two ratcheting effects. Notably, in \autoref{fig:x_vs_t_diffAlpha}(a) the temporal ratcheting effect when $\alpha=2$ is strong enough to reverse the direction of motion from downward to upward.

In what follows, we attempt to explicate the ratcheting behavior of the system (with more focus on the temporal ratchet) using a perturbation solution (in terms of $\epsilon\ll1$, i.e., weak nonlinear friction) to the problem. Note that we only consider the nonlinear component of the friction to be weak in \eqref{eq:governing_full}, which is consistent with the overdamped particle dynamics. We emphasize that such a temporal ratcheting phenomena has been reported previously in similar problems, including microfluidics \cite{Daniel2005Langmuir} and micromachining by solid-solid friction \cite{Fleishman2007,Aref2022PRE}. It is also worth mentioning that a working equation similar to \eqref{eq:governing_full} has been investigated before by Hashemi et al. \cite{Aref2022PRE} (cf. equation (1) in \cite{Aref2022PRE}). However, this prior work focused on a numerical study of the problem only. Our goal here is to provide a mathematical framework to \emph{theoretically} investigate such problems.

\section{Perturbation Solution}\label{sec:perturbation}
Here, we develop a perturbation solution to \eqref{eq:governing_full} to investigate the dynamics of the system at long times. It is readily shown that, even for a single mode excitation, a regular perturbation expansion yields unphysical secular terms and loses uniformity at times of order $\epsilon^{-1}$. (We demonstrate this for a simpler model problem in Appendix~\ref{sec:regular-perturbation}.) We therefore, use two-timing, and express the solution as a series in powers of $\epsilon$,
\begin{equation}
  \label{eq:perturbation}
  v(t)=v\p{0}\left(t_0,t_1\right)+\epsilon v\p{1}\left(t_0,t_1\right)+\epsilon^2 v\p{2}\left(t_0,t_1\right)+\dots
\end{equation}
where $t_0=t$ and $t_1=\epsilon t$ are the fast and slow time scales of the problem, respectively, so that $d/dt=\partial/\partial t_0+\epsilon\partial/\partial t_1$, and $v\p{i}$ are the solutions of different perturbation orders. We substitute \eqref{eq:perturbation} and definitions of $t_0$ and $t_1$ into \eqref{eq:governing_full}, and collect terms with equal powers of $\epsilon$, to obtain the system of equations for different perturbation orders.

The zeroth-order problem can be expressed as
\begin{equation}
\label{eq:zerothModeDE_full}
\frac{\partial v\p{0}}{\partial t_0}=\tfrac{1}{2}\left[\sin(t_0)+c\,\alpha^2\sin(\alpha t_0)\right]-kv\p{0}-w,
\end{equation}
which represents a balance between the particle acceleration and external forcing; that is, the weak nonlinear friction does not enter the leading order problem at the fast time scale. One can solve \eqref{eq:zerothModeDE_full} to obtain
\begin{align}
  \label{eq:sol_zeroth_full}
  v\p{0}(t_0,t_1)=&A(t_1)e^{-kt_0}+\beta\left[k\sin(t_0)-\cos(t_0)\right]\nonumber\\
  +&\gamma\left[\tfrac{k}{\alpha}\sin(\alpha t_0)-\cos(\alpha t_0)\right]-\frac{w}{k},
\end{align}
where $\beta=\tfrac{1}{2}/(1+k^2)$ and $\gamma=\tfrac{1}{2}c\,\alpha^3/(\alpha^2+k^2)$. Note that the constant of integration is a function of the slow time scale $t_1$. Equation \eqref{eq:sol_zeroth_full} suggests that $\langle v\p{0}\rangle=-w/k$ as $t\to\infty$, which is the linear response of the system to gravitational force.

To observe interesting nonlinear dynamics, we need to investigate higher perturbation orders. The first-order differential equation becomes
\begin{align}
  \label{eq:firstModeDE}
  \frac{\partial v\p{1}}{\partial t_0}=-&\frac{\partial v\p{0}}{\partial t_1}-kv\p{1}-\left[v\p{0}\right]^3\nonumber\\
  =-&\frac{dA}{dt_1}e^{-kt_0}-kv\p{1}\nonumber\\
  -\Big\{&Ae^{-kt_0}+\beta\left[k\sin(t_0)-\cos(t_0)\right]\nonumber\\
  +&\gamma\left[\tfrac{k}{\alpha}\sin(\alpha t_0)-\cos(\alpha t_0)\right]-\frac{w}{k}\Big\}^3.
\end{align}
Since the total resistance force is a monotonically increasing function of $v$, i.e., $d\left(kv+\epsilon v^3\right)/dv\ge 0$ (see \eqref{eq:governing_full}), the particle velocity cannot increase indefinitely, and hence, we require $\langle dv/dt\rangle=0$ at \emph{long times}. In other words, the harmonic solution has a constant average velocity as $t\to\infty$. Therefore, one needs to set the sum of the non-vanishing terms (upon time-averaging) on the RHS of \eqref{eq:firstModeDE} equal to zero, which provides an explicit expression for $\langle v\p{1}\rangle$:
\begin{align}
  \label{eq:first-order-contribution}
  \langle v\p{1}\rangle=-\tfrac{1}{k}\Big\langle\Big\{&\beta\left[k\sin(t_0)-\cos(t_0)\right]\nonumber\\
  +&\gamma\left[\tfrac{k}{\alpha}\sin(\alpha t_0)-\cos(\alpha t_0)\right]-\frac{w}{k}\Big\}^3\Big\rangle,
\end{align}

Note that the overall first-order time-average solution is $\langle v\p{0}\rangle+\epsilon \langle v\p{1}\rangle$. From \eqref{eq:sol_zeroth_full} the zeroth-order contribution is simply $\langle v\p{0}\rangle=-w/k$, the terminal velocity in the presence of linear drag. The first-order contribution given by \eqref{eq:first-order-contribution} is more complicated and in particular depends on the value of $\alpha$. \autoref{tab:v1_ave} lists $\langle v\p{1}\rangle$ formula from \eqref{eq:first-order-contribution} for different values of $\alpha$. This first-order perturbation accurately predicts the dynamics of the particle for sufficiently small values of $\epsilon$. For example, for $c=1,\,k=1,\,w=0.001,\,\epsilon=0.01,\,\alpha=2$, $\langle v\p{0}\rangle+\epsilon\langle v\p{1}\rangle=-6.11\times 10^{-4}$, which is close to the numerically obtained value of $-6.28\times 10^{-4}$ for the average velocity. More importantly, our first-order perturbation solution correctly captures the qualitative behavior of the system; it predicts a nonzero time-average velocity for $\alpha=2$ even in the absence of gravitational forces (i.e., $w=0$). However, it fails to predict the expected behavior ($\langle v\p{1}\rangle\neq 0$) for other values of $\alpha$ that yield non-antiperiodic excitations, such as $\alpha=\tfrac{3}{2}$. Note in \autoref{tab:v1_ave} that $\alpha=2$ is the only case that yields a time-average velocity contribution that is independent of $w$.

We argue that the contribution of other non-antiperiodic $\alpha$ values shows up at higher perturbation orders. To demonstrate this idea, we consider a simplified working equation, albeit less physically feasible for the original problem, in \autoref{sec:simpleModel}. Of course, our analysis can be extended to the original problem \eqref{eq:governing_full}; however, the algebraic complexity of the calculation would be greatly increased. Nonetheless, we expect that the conclusions gained below from analysis of the simplified problem will hold for the original problem.

\renewcommand{\arraystretch}{2}
\begin{table}[h]
\caption{$\langle v\p{1}\rangle$ formula for different values of $\alpha$.}
%% \begin{center}
  \begin{tabular}{c?{\tabthickness}c}
    $\alpha$ & $\langle v\p{1}\rangle$ \\
    \Xhline{\tabthickness}
    $1$            & $\frac{3\left(1+c\right)^2}{8k^2\left(1+k^2\right)}w\!+\!\frac{w^3}{k^4}$ \\
    $\tfrac{3}{2}$ & $\frac{3\left(4\left(9+4k^2\right)+81c^2\left(1+k^2\right)\right)}{32k^2\left(1+k^2\right)\left(9+4k^2\right)}w\!+\!\frac{w^3}{k^4}$ \\
    $\tfrac{5}{3}$ & $\frac{9\left(25+9k^2\right)+625c^2\left(1+k^2\right)}{24k^2\left(1+k^2\right)\left(25+9k^2\right)}w\!+\!\frac{w^3}{k^4}$ \\
    $\tfrac{19}{9}$ & $\frac{81\left(361+81k^2\right)+130321c^2\left(1+k^2\right)}{216k^2\left(1+k^2\right)\left(361+81k^2\right)}w\!+\!\frac{w^3}{k^4}$ \\
    $3$ & $\frac{3\left(9+k^2+81c^2\left(1+k^2\right)\right)}{8k^2\left(1+k^2\right)\left(9+k^2\right)}w\!+\!\frac{w^3}{k^4}$ \\
    $2$ & $\frac{3\left(4+k^2+16c^2\left(1+k^2\right)\right)}{8k^2\left(4+5k^2+k^4\right)}w\!+\!\frac{w^3}{k^4}\!+\!\frac{3c}{4k\left(1+k^2\right)^2\left(4+k^2\right)}$
  \end{tabular}
%% \end{center}
\label{tab:v1_ave}
\end{table}

\section{Simple Model}\label{sec:simpleModel}
In this section, we develop a high-order perturbation solution for a simplified version of \eqref{eq:governing_full},
\begin{equation}
  \label{eq:governing}
  \frac{dv}{dt}=-\frac{d^2f}{{dt}^2}-\epsilon v^3,
\end{equation}
where we have removed the linear drag term. Also, note that we are more interested in the temporal ratcheting effect and, hence, let $w=0$ to disable the spatial asymmetry. A variety of overdamped nonlinear systems can be reduced to \eqref{eq:governing}. For example, consider a solid-solid friction system in which a flat object is placed on a surface that vibrates laterally (cf. Hashemi et al. \cite{Aref2022PRE} for details). In the absence of a sticking regime (i.e., the object is always sliding on the surface), one obtains the dimensionless equation of motion as $dv/dt=-d^2\!f/{dt}^2-\lambda\,\mathrm{sgn}(v)$, where $\lambda$ is the dimensionless kinetic friction coefficient and $v$ is the object velocity relative to the surface. Clearly, this equation can be simplified to \eqref{eq:governing} using the same approximation to the signum function. Of course, in a more realistic solid-solid friction model, we need to account for the potential periods of time that the object sticks to the surface \cite{Aref2022PRE}.

We follow a similar procedure as that in \autoref{sec:perturbation} and consider a two-timing perturbation solution in terms of $\epsilon$. The zeroth-order differential equation becomes
\begin{equation}
  \label{eq:zeroModeDE}
  \frac{\partial v\p{0}}{\partial t_0}=\tfrac{1}{2}\left[\sin(t_0)+c\,\alpha^2\sin(\alpha t_0)\right],
\end{equation}
with general solutions
\begin{equation}
  \label{eq:sol_zeroth}
  v\p{0}(t_0,t_1)=-\tfrac{1}{2}\left[\cos(t_0)+c\,\alpha\,\cos(\alpha t_0)\right]+A(t_1).
\end{equation}
The first-order problem can then be written as
\begin{align}
  \label{eq:ode_first_compact}
  &\frac{\partial v\p{1}}{\partial t_0}=-\frac{\partial v\p{0}}{\partial t_1}-\left[v\p{0}\right]^3\nonumber\\
  &=-\frac{dA}{dt_1}+\left[\tfrac{1}{2}\left(\cos(t_0)+c\,\alpha\,\cos(\alpha t_0)\right)-A\right]^3.
\end{align}

Similar to our analysis in \autoref{sec:perturbation}, since $d\left(\epsilon v^3\right)/dv\ge 0$, the solution $v$ cannot increase without bound and we enforce $\langle dv/dt\rangle=0$. Hence, we set the sum of the non-vanishing terms (upon time-averaging) on the RHS of \eqref{eq:ode_first_compact} equal to zero.

Two distinctively different conditions happen for $\alpha=2$ and $\alpha\neq 2$. To see this, we need to expand the RHS of \eqref{eq:ode_first_compact} as
\begin{align}
  \label{eq:ode_first_expanded}
  &\frac{\partial v\p{1}}{\partial t_0}=\gamma-\tfrac{3}{8}A\left(1+c^2\alpha^2\right)-A^3-\frac{dA}{dt_1}\nonumber\\
  &+\tfrac{3}{8}c\,\alpha\,\cos^2(t_0)\cos(\alpha t_0)\!-\!\tfrac{3}{2}A\,c\,\alpha\,\cos(t_0)\cos(\alpha t_0),
\end{align}
where, $\gamma$ is a collection of cosine terms and $\langle\gamma\rangle=0$. If $\alpha\neq2$, we require
\begin{equation}
  \label{eq:A_alphaneq2}
  \frac{dA}{dt_1}=-\tfrac{3}{8}\left(1+c^2\alpha^2\right)A-A^3,
\end{equation}
to ensure that the time-average of the RHS of \eqref{eq:ode_first_expanded} over the period of the solution goes to zero. One can solve \eqref{eq:A_alphaneq2} to find
\begin{equation}
  \label{eq:Asolve_alphaneq2}
  A=\pm\sqrt{\frac{\beta}{Ke^{2\beta t_1}-1}},
\end{equation}
for some constant $K$ and $\beta=\tfrac{3}{8}\left(1+c^2\alpha^2\right)$. Clearly, $A\to0$ as $t_1$, and hence $t$, go to infinity. Therefore, at long times, when the harmonic solution is achieved, there will be no net response. Alternatively, note that the only real fixed-point of \eqref{eq:A_alphaneq2} is $A^*=0$, which points to the same conclusions. Note that $A$ does not change sign from its initial value $A(0)=\tfrac{1}{2}(1+c\,\alpha)$. So the correct sign in \eqref{eq:Asolve_alphaneq2} is determined by the sign of $\tfrac{1}{2}(1+c\,\alpha)$. But, perhaps, the sign is insignificant here as $A$ goes to zero regardless of the initial conditions for $\alpha\neq2$. When $\alpha=1$, the last term on the RHS of \eqref{eq:ode_first_expanded} gives rise to another steady term $-\tfrac{3}{4}A\,c$, and we need
\begin{equation}
  \label{eq:A_alphaeq1}
  \frac{dA}{dt_1}=-\tfrac{3}{8}(1+c)^2A-A^3,
\end{equation}
which again has a zero fixed-point $A^*=0$.

For $\alpha=2$, the term $\tfrac{3}{8}c\,\alpha\cos^2(t_0)\cos(\alpha t_0)$ in \eqref{eq:ode_first_expanded} becomes $\tfrac{3}{16}c+\tfrac{3}{16}c\left(\cos(4t_0)+2\cos(2t_0)\right)$, and therefore, we enforce
\begin{equation}
  \label{eq:A_alphaeq2}
  \frac{dA}{dt_1}=\tfrac{3}{16}c-\tfrac{3}{8}\left(1+4c^2\right)A-A^3,
\end{equation}
an Abel differential equation of the first kind \cite{HandbookDE}, with one real, \emph{nonzero}, fixed point given by
\begin{equation}
  \label{eq:Astar_alphaeq2}
  A^*=\frac{2^{\tfrac{2}{3}}\left(1+4c^2\right)-2^{\tfrac{1}{3}}\zeta^{\tfrac{2}{3}}}{4\zeta^{\tfrac{1}{3}}},
\end{equation}
where $\zeta=-3c+\sqrt{2+33c^2+96c^4+128c^6}$. This nonzero fixed point is, to leading order, the net response of the system when $\alpha=2$.

\begin{figure}[t]
  \centering
  \setlength{\belowcaptionskip}{-10pt}
  \includegraphics{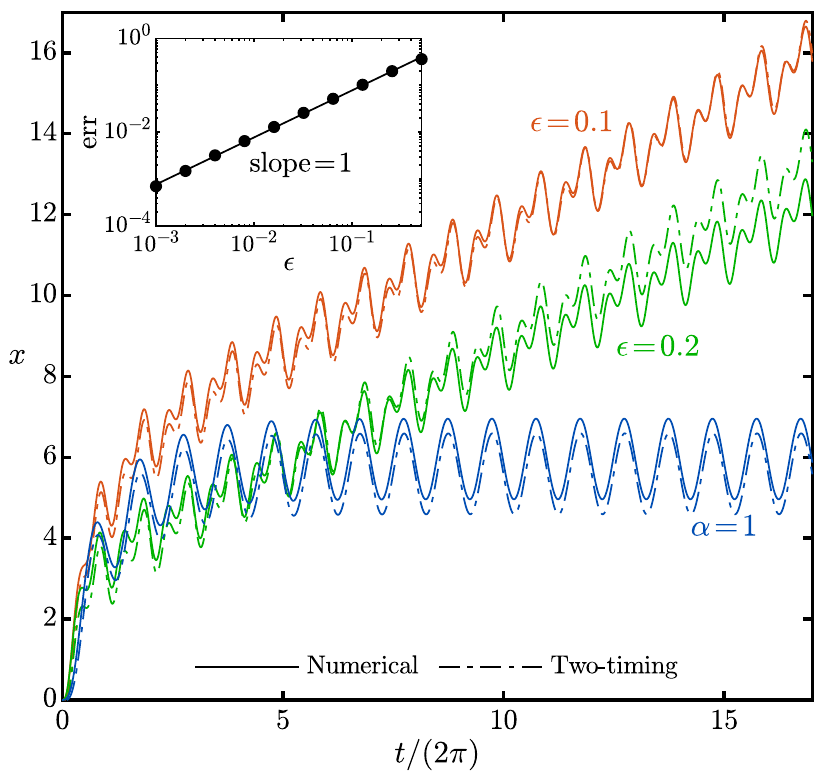}
  \caption{Numerical results and two-timing perturbation solutions to $x(t)=\int_0^tv(s)ds$ from \eqref{eq:governing} versus time, for $\alpha=1$ ($\epsilon=0.1$) and $\alpha=2$ (and $\epsilon=0.1,\,0.2$). The two-timing solution is $x\p{0}(t)$, obtained by integrating \eqref{eq:sol_zeroth}, with the slowly varying $A(t_1)$ given by the differential equations \eqref{eq:A_alphaeq1} and \eqref{eq:A_alphaeq2} for $\alpha=1$ and $2$, respectively. The inset shows normalized norm of difference between the numerical ($v(t)$) and two-timing ($v\p{0}(t)$) solutions to the harmonic solution $v$, i.e., $\text{err}=\lVert v(t)-v\p{0}(t)\rVert/\lVert v(t)\rVert$ (over the solution period), versus $\epsilon$. Parameter: $c=1$}
  \label{fig:NAcompare}
\end{figure}

\autoref{fig:NAcompare} shows a comparison between the numerical and two-timing solutions to $x(t)=\int_0^tv(s)ds$ for $\alpha=2$ and different $\epsilon$, along with a reference case of $\alpha=1$. To obtain $A(t_1)$, and subsequently, the two-timing solution $v\p{0}(t)$ (and $x\p{0}(t)$ by integration) versus time, we numerically solve the initial value problems \eqref{eq:A_alphaeq1} and \eqref{eq:A_alphaeq2} for $\alpha=1$ and $2$, respectively. Our results show that the two-timing solution agrees well with the numerical solutions for small $\epsilon$ values. (The inset in \autoref{fig:NAcompare} illustrates the error measure $\lVert v(t)-v\p{0}(t)\rVert/\lVert v(t)\rVert$ for one period of the harmonic solutions.) The case of $\alpha=1$ (with $\epsilon=0.1$), for which we expect no net response, also indicates good agreements between the two solutions.

\begin{figure}[t]
  \centering
  \setlength{\belowcaptionskip}{-10pt}
  \includegraphics{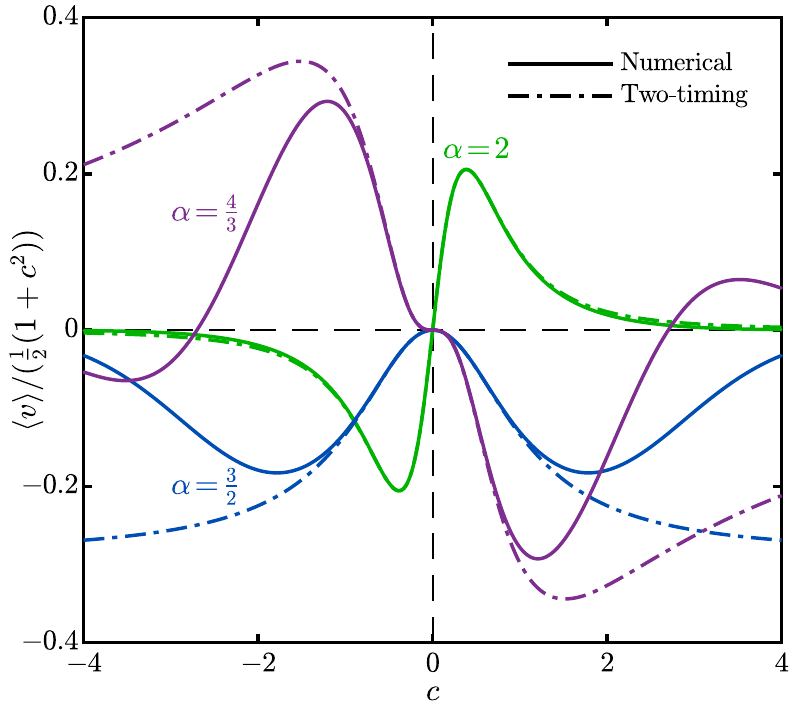}
  \caption{Numerical results and two-timing perturbation solutions to the harmonic time-average solution $\langle v(t)\rangle$ from \eqref{eq:governing} versus $c$, for $\alpha=2,\,\tfrac{3}{2},\,\tfrac{4}{3}$. The two-timing solutions are given by $A^*$ (\eqref{eq:Astar_alphaeq2}) for $\alpha=2$, and $\epsilon^2C^*$ for $\alpha=\tfrac{3}{2}$ and $\tfrac{4}{3}$ (\eqref{eq:Cstar_alphaeq1.5} and \eqref{eq:Cstar_alphaeq1.333}, respectively). For visualization purposes, the data for $\alpha=\tfrac{3}{2}$ and $\tfrac{4}{3}$ are multiplied by $10^2$ and $10^3$, respectively. Parameters: $\epsilon=0.05$.}
  \label{fig:cEffect}
\end{figure}

Interestingly, similar to the original model problem in \autoref{sec:perturbation}, the above first-order two-timing perturbation solution predicts the ratchet effect only for $\alpha=2$, and not for the other non-antiperiodic waveforms, such as those for $\alpha=\tfrac{3}{2}$ and $\tfrac{4}{3}$. To see these yet weaker nonlinear phenomena, we need to investigate higher perturbation orders. In Appendix~\ref{sec:derivation_higherorder} we show that one needs to go up to the third perturbation order to find information about the ratchet effects at the next $\alpha$ values with the strongest ratchet effect. Our analysis suggests that $\langle v\rangle\to\epsilon^2C^*$ as $t\to\infty$ (harmonic solution), where $C^*$ is given by \eqref{eq:Cstar_alphaeq1.5} and \eqref{eq:Cstar_alphaeq1.333} for $\alpha=\tfrac{3}{2}$ and $\tfrac{4}{3}$, respectively.

In \autoref{fig:cEffect}, we compare the numerical and two-timing solutions to the time-average solution $\langle v(t)\rangle$ for $\alpha=2$ ($A^*$ from \eqref{eq:Astar_alphaeq2}), $\tfrac{3}{2}$ ($\epsilon^2C^*$ from \eqref{eq:Cstar_alphaeq1.5}), and $\tfrac{4}{3}$ ($\epsilon^2C^*$ from \eqref{eq:Cstar_alphaeq1.333}). Let us first discuss the observed, intriguing, effects of $c$. Recall, $c$ is the relative magnitude of two modes in the excitation. For $\alpha=2$ and $\tfrac{4}{3}$, it is the sign of $c$ that determines the direction of motion, while when $\alpha=\tfrac{3}{2}$, $\langle v\rangle$ is always negative. Such a qualitative difference between the cases for $\alpha=\tfrac{3}{2}$ and $\frac{4}{3}$ is counterintuitive, given that both their net responses are induced by the third perturbation order. Additionally, note that the effect of $c$ is non-monotonic for all of the $\alpha$ values. When $c=0$, the waveform \eqref{eq:f} reduces to a single-mode wave oscillating as $\sin(t)$, which yields no net response. Similarly, for large values of $c$, the waveform mimics a single-mode oscillator $\sin(\alpha t)$, which again leads to a zero net response. (Of course, the latter effect is exaggerated in \autoref{fig:cEffect} due to normalization by $\tfrac{1}{2}\left(1+c^2\right)$.) Consequently, there is a $c$ for which the ratchet effect is maximized. For the representative results shown in \autoref{fig:cEffect}, when $\alpha=2$, this maximum value occurs at $c\approx 0.38$ and is equal to $\langle v\rangle\approx 0.2\times\tfrac{1}{2}\left(1+c^2\right)\approx 0.11$, which is $\approx1.4\,\mathrm{mm/s}$ for a vibration with $\ell=0.5\,\mathrm{mm}$ at $4\,\mathrm{Hz}$. Another important observation in \autoref{fig:cEffect} is the relatively inaccurate predictions of the two-timing method at large values of $\mid\!c\!\mid$ (particularly for $\alpha=\tfrac{3}{2}$ and $\tfrac{4}{3}$). However, this is an expected behavior, given that our perturbation solution is built upon the assumption that $f(t)$, and therefore, $c=O(1)$.

\section{Conclusions}
The present analysis points up further interesting questions. For instance, for a given $\alpha$, what perturbation orders contribute to the temporal ratchet effect? An answer to this question can help explain the relative magnitude of the net response for different $\alpha$ values. The two-timing methodology provided here can be employed for the study of temporal ratchets in other, more complicated, systems. Specific future works include: (i) an experimental analysis of the current problem; and (ii) developing a two-timing perturbation solution to the Poisson-Nernst-Planck equations to study the ratchet asymmetric rectified electric field \cite{Aref2022PRE}. Finally, our work elucidates that temporal ratchets in weakly nonlinear systems arise from the cumulative effect of instantaneously small nonlinear disturbances. These disturbances accumulate to generate a leading order drift at long times. It has been shown previously \cite{Aref2022PRE} that temporal ratchets are maximal for $\alpha=2$ excitations: our work explains this finding, since such excitations are shown to produce net drift at fastest slow time-scale. In principle, while net drift could occur for any non-antiperiodic signal, it may occur on such a long-time scale to practically impossible to observe in experiment.

\section*{Acknowledgments}
This material is partially based upon work conducted during the 2022 AM-SURE program at the Courant Institute, supported by the National Science Foundation under Grant No. RTG/DMS-1646339. We thank Leif Ristroph for helpful discussions regarding future experimental investigation of the problem.

\section*{Statements and Declarations}
\noindent{\textbf{Competing Interests.}} There are no conﬂicts to declare.\\
\noindent{\textbf{Author contributions.}} A.H. performed the computational study and wrote the manuscript; A.H., E.T.G., and A.S.K. developed the perturbation solution and conducted the analytical study of the problem; A.H. and A.S.K. envisioned and led the project.\\
\noindent{\textbf{Data availability.}} The data that support the findings of this study are available upon request from the corresponding author [A.H.].

\begin{appendices}
\section{Regular perturbation}\label{sec:regular-perturbation}
In this section, we show that a regular perturbation expansion fails to capture the physics of the problem at long times, even for a single-mode vibration. We consider the model given by \eqref{eq:governing}. Let $f=\sin(t)$ and $v(t)\approx v\p{0}(t)+\epsilon v\p{1}(t)+O(\epsilon^2)$, which upon substituting into \eqref{eq:governing}, and solving for $v\p{0}$ yield
\begin{equation}
  v\p{0}=-\cos(t)+A.
\end{equation}
Using the initial condition $v(t=0)=v\p{0}(t)+\epsilon v\p{1}(t)$, we conclude that $v\p{0}=0$ and $v\p{1}=0$, which we use to find the constant $A=1$. The first-order differential equation is therefore $dv\p{1}/dt=(\cos(t)-1)^3$, with a general solution
\begin{equation}
  v\p{1}=\tfrac{1}{12}\left[\sin(3t)-9\sin(2t)+45\sin(t)\right]-\tfrac{5}{2}t+B,
\end{equation}
where the initial condition $v\p{1}(0)=0$ gives $B=0$. Hence, the overall, first-order, solution is
\begin{align}
  v(t)&=1-\cos(t)-\tfrac{5}{2}\epsilon t\nonumber\\
  &+\tfrac{1}{12}\epsilon\left[\sin(3t)-9\sin(2t)+45\sin(t)\right].
\end{align}
This solution predicts, incorrectly, that the time-average solution increases indefinitely with time. Also, it loses uniformity at long times ($t=O(\epsilon^{-1})$), suggesting that the system behavior is governed by a fast, $t$, and a slow, $\epsilon t$, time scales.  

\section{Higher perturbation orders}\label{sec:derivation_higherorder}
Here, for model problem \eqref{eq:governing}, we demonstrate how the higher perturbation orders contribute to the observed ratchet effect for $\alpha=\tfrac{3}{2}$ and similarly $\alpha=\tfrac{4}{3}$. These are the next two $\alpha$ values after $\alpha=2$ with the strongest response.

We focus on finding the \emph{long-time} solution to the time-average response, i.e., $t\to\infty$. Under such conditions, $A(t_1)$ given by \eqref{eq:Asolve_alphaneq2} goes to zero, and so does $dA/dt_1$ (when $\alpha\neq 2$). Therefore, the zeroth-order solution simplifies to
\begin{equation}
  \label{eq:v0-appendix}
  v\p{0}=-\tfrac{1}{2}\left[\cos(t_0)+\alpha c\cos(\alpha t_0)\right].
\end{equation}
Likewise, the first-order differential equation \eqref{eq:ode_first_compact} reduces to
\begin{equation}
  \label{eq:ode-v1-appendix}
  \frac{\partial v\p{1}}{\partial t_0}=\tfrac{1}{8}\left[\cos(t_0)+\alpha c\cos(\alpha t_0)\right]^3,
\end{equation}
with a general solution of the form
\begin{equation}
  \label{eq:v1-appendix}
  v\p{1}=\{\text{sine terms}\}(t_0)+B(t_1).
\end{equation}
Here, $B(t_1)$ is a constant of integration and is a function of the slow time scale, and $\{\text{sine terms}\}(t_0)$ is a collection of sine terms.

The second-order differential equation has the form
\begin{equation}
  \label{eq:ode-v2-appendix}
  \frac{\partial v\p{2}}{\partial t_0}\!=\!-3\left[v\p{0}\right]^2\!\!\!v\p{1}\!-\!\frac{\partial v\p{1}}{\partial t_1}\!=\!-3\left[v\p{0}\right]^2\!\!\!v\p{1}\!-\!\frac{dB}{dt_1}.
\end{equation}
It is rather straightforward to show that the product $\left[v\p{0}\right]^2v\p{1}$ does not introduce any secular term to the RHS of \eqref{eq:ode-v2-appendix}. As a result, enforcing the time-average of the RHS to be zero, yields an ordinary differential equation (ODE) for $B(t_1)$ with a zero fixed-point. When $\alpha=\tfrac{3}{2}$, this ODE can be expressed as $dB/dt_1=-\tfrac{3}{32}(4+9c^2)B$. Clearly, $B\to0$ as $t\to\infty$. So, similar to $A(t_1)$, we set $B(t_1)=0$ in our analysis of the long-time behavior. The general solution for $v\p{2}$ becomes
\begin{equation}
  \label{eq:v2-appendix}
  v\p{2}=\{\text{cosine terms}\}(t_0)+C(t_1),
\end{equation}
where $\{\text{cosine terms}\}(t_0)$ is a collection of cosine terms and $C(t_1)$ is the constant of integration, which is to be determined.

The third-order differential equation is
\begin{equation}
  \label{eq:ode-v3-appendix}
  \frac{\partial v\p{3}}{\partial t_0}=-3\left(\left[v\p{0}\right]^2\!\!\!v\p{2}+\left[v\p{1}\right]^2\!\!\!v\p{0}\right)-\frac{\partial v\p{2}}{\partial t_1}.
\end{equation}
Again, we require that the sum of the nonzero time-average terms on the RHS to vanish, which leads to an ODE for $C(t_1)$. For $\alpha=\tfrac{3}{2}$, the fixed point $C^*$ of this ODE is
\begin{equation}
  \label{eq:Cstar_alphaeq1.5}
  C^*=-\frac{3c^2\left(33873+98935c^2\right)}{57344\left(4+9c^2\right)}.
\end{equation}
Similarly, for $\alpha=\tfrac{4}{3}$, we have
\begin{equation}
  \label{eq:Cstar_alphaeq1.333}
  C^*=-\frac{34349c^3}{11520\left(9+16c^2\right)}.
\end{equation}
Hence, the induced time-average solution for $\alpha=\tfrac{3}{2}$ and $\tfrac{4}{3}$ can be approximated as $\epsilon^2C^*$, where $C^*$ is given by \eqref{eq:Cstar_alphaeq1.5} and \eqref{eq:Cstar_alphaeq1.333}, respectively.

%% \begin{align*}
%%   &\gamma\!=\!\tfrac{3}{2}A^2\left(\cos(t_0)\!+\!c\,\alpha\cos(\alpha t_0)\right)\!-\!\tfrac{3}{8}A\left(\cos(2t_0)\!+\!c^2\alpha^2\cos(2\alpha t_0)\right)\nonumber\\
%%   &+\tfrac{1}{8}\left(\cos^3(t_0)+c^3\alpha^3\cos^3(\alpha t_0)+3c^2\alpha^2\cos^2(\alpha t_0)\cos(t_0)\right),
%% \end{align*}

%% , e.g., for $\alpha=\tfrac{3}{2}$, $\{\text{sine terms}\}(t_0)=$
%% \begin{align*}
%%     &\tfrac{1}{10752}\Big[3024c\,\sin(\tfrac{1}{2}t_0)\!+\!504\left(2+9c^2\right)\sin(t_0)\\
%%     &\!+\!c\left(2016\!+\!2268c^2\right)\sin(\tfrac{3}{2}t_0)\!+\!1134c^2\sin(2t_0)\!+\!112\sin(3t_0)\\
%%     &\!+\!432c\,\sin(\tfrac{7}{2}t_0)\!+\!567c^2\sin(4t_0)\!+\!252c^3\sin(\tfrac{9}{2}t_0)\Big].  
%% \end{align*}

\end{appendices}

%%===========================================================================================%%
%% If you are submitting to one of the Nature Portfolio journals, using the eJP submission   %%
%% system, please include the references within the manuscript file itself. You may do this  %%
%% by copying the reference list from your .bbl file, paste it into the main manuscript .tex %%
%% file, and delete the associated \verb+\bibliography+ commands.                            %%
%%===========================================================================================%%

%% \bibliography{../REFERENCES}% common bib file
%% if required, the content of .bbl file can be included here once bbl is generated
%%\input sn-article.bbl

%% BioMed_Central_Bib_Style_v1.01

\end{document}